\documentclass[
twocolumn]{aastex631}

\usepackage{apjfonts}

\DeclareSymbolFont{cmletters}{OML}{cmm}{m}{it}
\DeclareMathSymbol{v}{\mathalpha}{cmletters}{"76}

\newcommand{\hammer}{H-AMR}
\newcommand{\phibh}{\phi_{\rm BH}}
\shorttitle{Truncated Accretion Disks}
\shortauthors{Liska et al.}
\graphicspath{{./}{figures/}}
\begin{document}

\title{Formation of Magnetically Truncated Accretion Disks in 3D Radiation-Transport Two-Temperature GRMHD Simulations}

\correspondingauthor{Matthew Liska}
\email{matthew.liska@cfa.harvard.edu}
\author{M.T.P. Liska}
\affiliation{Institute for Theory and Computation, Harvard University, 60 Garden Street, Cambridge, MA 02138, USA}
\author{G. Musoke}
\affiliation{Anton Pannekoek Institute for Astronomy, University of Amsterdam, Science Park 904, 1098 XH Amsterdam, The Netherlands}
\author{A. Tchekhovskoy}
\affiliation{Center for Interdisciplinary Exploration \& Research in Astrophysics (CIERA), Physics \& Astronomy, Northwestern University, Evanston, IL 60202, USA}
\author{O. Porth}
\affiliation{Anton Pannekoek Institute for Astronomy, University of Amsterdam, Science Park 904, 1098 XH Amsterdam, The Netherlands}
\author{A. M. Beloborodov}
\affiliation{Physics Department and Columbia Astrophysics Laboratory, Columbia University, 538 West 120th Street, New York, NY 10027, USA}
\affiliation{Max Planck Institute for Astrophysics, Karl-Schwarzschild-Str. 1, D-85741, Garching, Germany}

%% Note that the \and command from previous versions of AASTeX is now
%% depreciated in this version as it is no longer necessary. AASTeX 
%% automatically takes care of all commas and "and"s between authors names.

%% AASTeX 6.31 has the new \collaboration and \nocollaboration commands to
%% provide the collaboration status of a group of authors. These commands 
%% can be used either before or after the list of corresponding authors. The
%% argument for \collaboration is the collaboration identifier. Authors are
%% encouraged to surround collaboration identifiers with ()s. The 
%% \nocollaboration command takes no argument and exists to indicate that
%% the nearby authors are not part of surrounding collaborations.

%% Mark off the abstract in the ``abstract'' environment. 
\begin{abstract}
Multi-wavelength observations suggest that the accretion disk in the hard and intermediate states of X-ray binaries (XRBs) and active galactic nuclei (AGN) transitions from a cold, thin disk at large distances into a hot, thick flow close to the black hole. However, the formation, structure and dynamics of such truncated disks are poorly constrained due to the complexity of the thermodynamic, magnetic, and radiative processes involved. We present the first radiation-transport two-temperature general relativistic magnetohydrodynamic (GRMHD) simulations of truncated disks radiating at $\sim 35\%$ of the Eddington luminosity with and without large-scale poloidal magnetic flux. We demonstrate that when a geometrically-thin accretion disk is threaded by large-scale net poloidal magnetic flux, it self-consistently transitions at small radii into a two-phase medium of cold gas clumps floating through a hot, magnetically dominated corona. This transition occurs at a well-defined truncation radius determined by the distance out to which the disk is saturated with magnetic flux. The average ion and electron temperatures in the semi-opaque corona reach, respectively, $T_{i} \gtrsim 10^{10}K$ and $T_{e}\gtrsim 5 \times 10^8K$. The system produces radiation, powerful collimated jets and broader winds at the total energy efficiency exceeding $90\%$, the highest ever energy extraction efficiency from a spinning black hole by a radiatively efficient flow in a GRMHD simulation. This is consistent with jetted ejections observed during XRB outbursts. The two-phase medium may naturally lead to broadened iron line emission observed in the hard state.
\end{abstract}

%% Keywords should appear after the \end{abstract} command. 
%% The AAS Journals now uses Unified Astronomy Thesaurus concepts:
%% https://astrothesaurus.org
%% You will be asked to selected these concepts during the submission process
%% but this old "keyword" functionality is maintained in case authors want
%% to include these concepts in their preprints.
%\keywords{black hole physics (251) --- Ultraviolet astronomy(1736) --- History of astronomy(1868) --- Interdisciplinary astronomy(804)}

\section{Introduction}
\label{sec:Intro}
Observations spanning the radio to gamma-ray frequency range show that black hole (BH) X-ray binaries (XRBs) cycle through different spectral states of accretion over the course of months to years (e.g \citealt{Esin1997, Remillard2006}). BH XRBs spend most of their time in a dim, quiescent state. However, from time to time they go into outburst during which they transition into orders of magnitude more luminous spectral states some of which are associated with powerful collimated outflows, or jets. During such spectral state transitions, the spectrum of the source changes from a power-law--like `hard' spectrum to a black-body--like `soft' spectrum.  During the outbursts, the luminosity $L$ can reach a substantial fraction of the Eddington limit, $L_{Edd}$, at which the radiation forces become comparable with gravity. In some extreme cases, such as in extra-galactic ultra-luminous X-Ray sources (ULXs) and in the galactic XRB SS433, the inferred bolometric luminosity can even exceed the Eddington limit (e.g. \citealt{Fabrika2004, Fuchs2006,  Begelman2006, Middleton20201}), which can potentially be a sign of super-Eddington accretion \citep{Sadowski2015_Jet}. In fact, it is conceivable that most black hole growth occurs during outbursts when the disk is accreting at or above the Eddington limit (e.g. \citealt{volonteri07}). In particular, the `luminous-hard state' is very interesting since it is associated with the most powerful transient jets in XRBs (e.g., \citealt{Fender2004}) and Fanaroff-Riley type 2 (FRII, \citealt{fr74}) jets in luminous AGN (e.g. \citealt{Tchekhovskoy2015,2020ARA&A..58..407D}).  

Accretion disks are generally described by either a geometrically-thin, optically-thick disk \citep{ss73, Novikov1973} or a geometrically-thick, optically-thin radiatively inefficient accretion flow (RIAF) such as an advection dominated accretion flow (ADAF, see \citealt{Narayan1994}). In a geometrically-thin disk the density and optical depth are high. The balance between the radiative cooling and turbulent dissipation rates determines the geometric thickness, or the scaleheight, of the disk. Such geometrically-thin disk models can explain thermal emission in the high-soft state. In a RIAF, the density is so low that the plasma decouples into a two-temperature fluid, in which the Coulomb collisions between the ions and electrons become too rare to equilibrate the temperatures of the two species. This thermal decoupling, along with the preferential heating of the ions, prevents the disk from radiating a dynamically significant amount of dissipated energy. RIAFs can explain emission in the quiescent and hard spectral states \citep[e.g.][and references therein]{AbramowiczFragile2013}.

When a quiescent black hole accretion system goes into an outburst, it transitions into a hybrid state that features both the thermal `soft' and non-thermal `hard' components in its spectrum. This suggests the presence of both a standard accretion disk and a RIAF (e.g. \citealt{Remillard2006}). In the literature, `truncated' disk models form a popular paradigm describing such hybrid states  (e.g. \citealt{Esin1997, Ferreira2006,  Done2007, Marcel2018A}), including their quasi-periodic variability (e.g. \citealt{Ingram2009}). In these truncated disk models, the outer part of the disk is geometrically-thin while the inner part is geometrically-thick, hot, and described by an RIAF. We refer to the inner, hot RIAF--like part of the system as the corona. The nature of the corona is actively debated, and its origin and geometry -- often represented as a hot `lamppost' -- are not definitively known. 
The corona can Comptonize its own cyclo-synchrotron emission in addition to thermal emission from a thin accretion disk (e.g. \citealt{Wardzinsky2000, DelSanto2013}). It is important to note that the power-law emission attributed to the corona can make up a significant fraction of the total X-ray luminosity in BHXRBs and AGN and hence in such cases the hot coronal gas is likely to be also dynamically important \citep[e.g.][and references therein]{Reynolds_Fabian1997,Fukumura_Kazanas2007,Wilkins_Fabian2012}.

A major point of contention concerns the behavior of the truncation radius, $r_{\rm tr}$, which marks the transition from the outer geometrically-thin, radiatively efficient disk to the inner geometrically-thick, radiatively inefficient corona. It is not clear if the softening of the spectrum during a state transition is primarily caused by the truncation radius moving inwards \citep{Fender2004} or the corona shrinking in size (e.g. \citealt{Kara2019}), and what roles Compton-upscattered X-ray emission from the base of the jet (e.g. \citealt{Markoff2005}) and/or relativistic winds \citep{Beloborodov1999} play. In addition, the presence of broadened iron lines (e.g. \citealt{Reis2010}), which are typically associated with a reservoir of cold optically-thick gas near the innermost stable circular orbit (ISCO), in the low-hard state of several BHXRB systems is inconsistent with the standard picture of a truncated disk, which does not predict the presence of any cold gas within the truncation radius (e.g. \citealt{Done2006, Done2007}). 

Simulating truncated accretion disks remains extremely challenging because both radiation and thermal decoupling between ions and electrons is expected to play a prominent role. Current MHD simulations only partly addressed disk truncation by simplifying the (radiative) physics with an ad-hoc cooling function which led to a smooth transition \citep{Hogg2017, Hogg2018}. Here, we use the state-of-the-art radiation-transport two-temperature GRMHD simulations to demonstrate that truncated accretion disks naturally form in the presence of large-scale net poloidal magnetic flux. In Sections \ref{sec:Numerics} and \ref{sec:Physical_Setup} we describe our GRMHD code \hammer{} and the initial conditions, before  presenting our results and concluding in Sections \ref{sec:Results} and \ref{sec:conclusion}.

\section{Numerical Setup}
\label{sec:Numerics}
We use for our simulations the graphical processing unit (GPU) accelerated GRMHD code \hammer{} (\citealt{Liska2018A, Liska2020}; see \citealt{Porth2019} for code comparison). H-AMR is triple level parallelized using a hybrid CUDA-OpenMP-MPI framework that solves the equations of GRMHD in the Kerr-Schild foliation. Magnetic fields are evolved using a staggered grid approached based on \citet{Gardiner2005}. The spatial reconstruction of primitive variables from cell centers to cell faces is done using a third order accurate PPM method \citep{Collela1984}, which leads to second order overall spatial accuracy (cross-dimensional terms are not taken into account). Integration in time is performed using the second order IMEX2A scheme described in \citet{Mckinney2013}. \hammer{} features a flexible adaptive mesh refinement (AMR) framework that allows for the refinement both in space and time. Namely, our local adaptive time-stepping approach allows different blocks to have different time steps even at the same refinement level \citep{Liska2020}.

\begin{figure}
    \centering
    \includegraphics[clip,trim=0.0cm 0.0cm 0.0cm 0.0cm,width=\columnwidth]{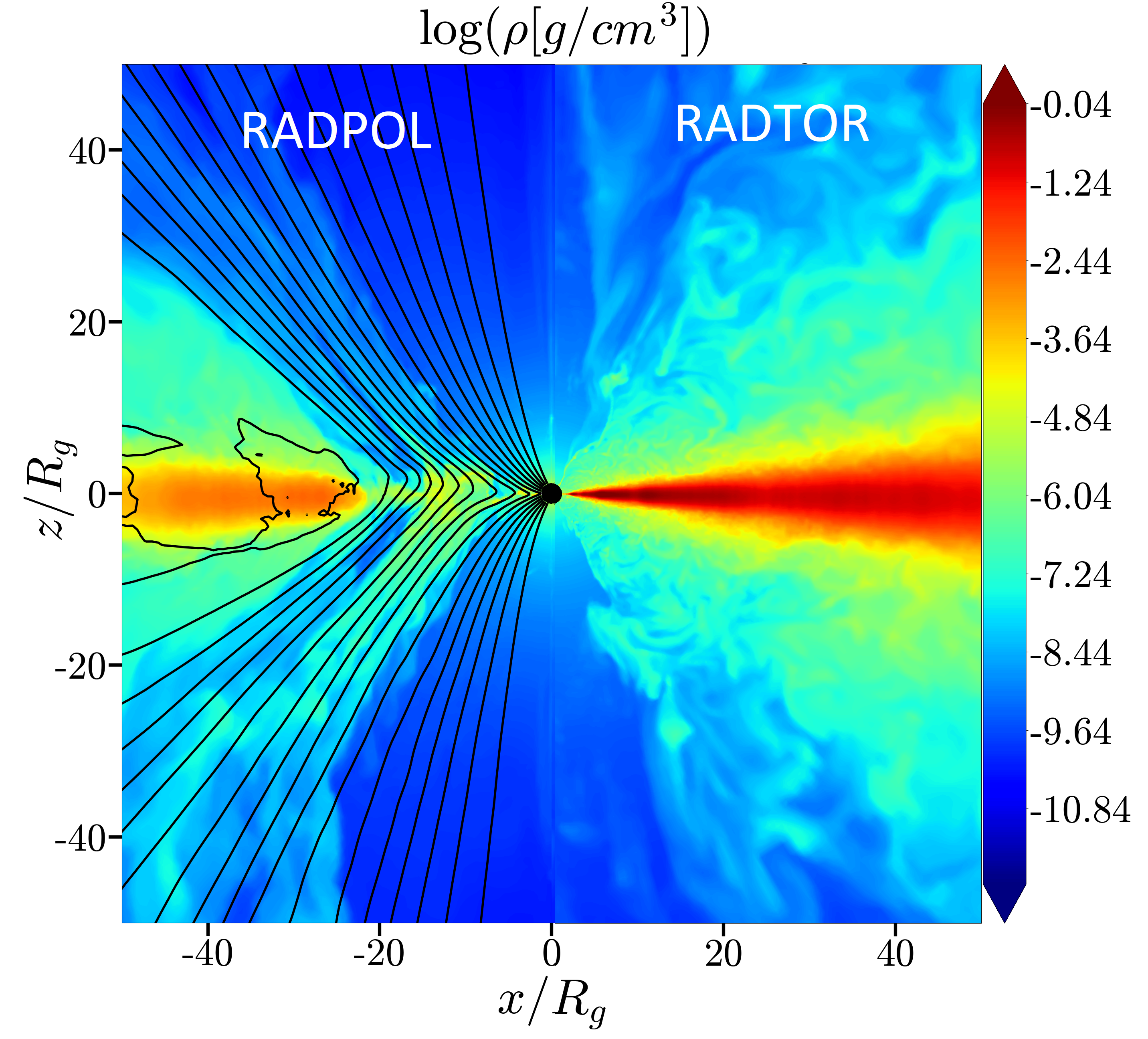}
    \vspace{-10pt}
    \caption{A transverse slice through density with magnetic field lines shown in black for our model with poloidal magnetic flux (RADPOL, left) and model with toroidal magnetic flux (RADTOR, right) right before enabling the radiation transport. While the RADTOR simulation forms a thin accretion disk with a constant scaleheight, the inner disk in the RADPOL simulation enters the MAD state and accretion proceeds through non-axisymmetric RTI modes.}
    \label{fig:init}
\end{figure}

We have recently implemented on-the-fly radiation transport into \hammer{} using an M1 (two-moment) closure, following the implementations in Koral \citep{Scadowski2013, Sadowski2017} and HARMRAD \citep{Mckinney2013}. We include the energy-averaged bound-free, free-free, and cyclo-Synchrotron absorption ($\kappa_{abs}$) and emission ($\kappa_{em}$) opacities in addition to the electrons scattering ($\kappa_{es}$) opacity as given in \citet{Mckinney2017}. We use grey-opacities energy-averaged over a diluted blackbody spectrum \citep{Mckinney2017} to provide as accurate as possible opacities within a single energy M1-framework. In addition, we assume that only electrons contribute and set $\kappa_{ion}=0$. This is valid since in high-density regions where bound-free opacities ($\kappa_{bf} \propto \rho$) become dominant, Coulomb collisions typically equilibrate the plasma to a single-temperature. 

\begin{figure}
    \centering
    \includegraphics[clip,trim=0.0cm 0.0cm 0.0cm 0.0cm,width=\columnwidth]{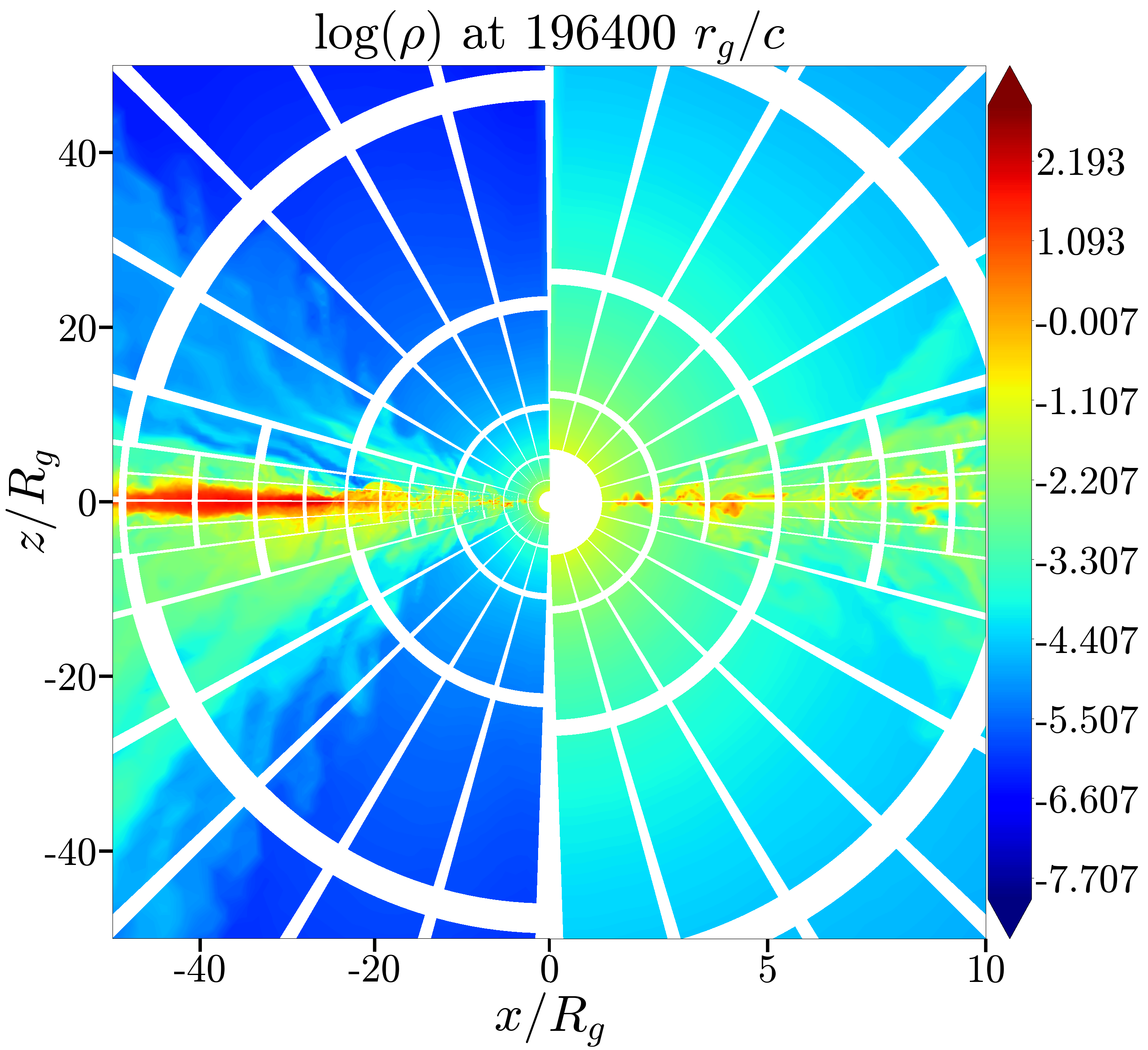}
    \vspace{-10pt}
    \caption{A transverse slice through RADPOL. The right hemisphere shows a zoom-in of the left hemisphere. The mesh-block boundaries are illustrated using white lines. Each mesh-block has $N_{r} \times N_{\theta} \times N_{\phi} = 36 \times 36 \times 102$ cells.}
    \label{fig:gridplot}
\end{figure}

The M1 closure works well in the optically-thick disk body and outside of the thin disk where the radiation moves predominantly in one direction (perpendicular to the disk surface). Since this is not always valid in the corona we further discuss the physical implications of the M1 approximation on our results in section \ref{sec:conclusion}.  To approximate Compton cooling of the corona we implemented both blackbody and photon number conserving Comptonization in \hammer{} \citep{Sadowski2015}. By evolving the photon number ($n_{rad}$), a more accurate radiation temperature can be calculated in the optically thin corona ($T_{rad} \sim (E_{rad} / (2.7 k_b n_{rad})$ instead of $T_{rad}=(E_{rad}/a)^{1/4}$). This increases the accuracy of the temperature dependent emission, absorption and Compton scattering coefficients in the plasma, and might play an important effect when the radiation spectrum deviates from a blackbody such as in an optically thin corona.

\begin{figure*}
\begin{center}
\includegraphics[width=1.0\linewidth,trim=0mm 0mm 0mm 0,clip]{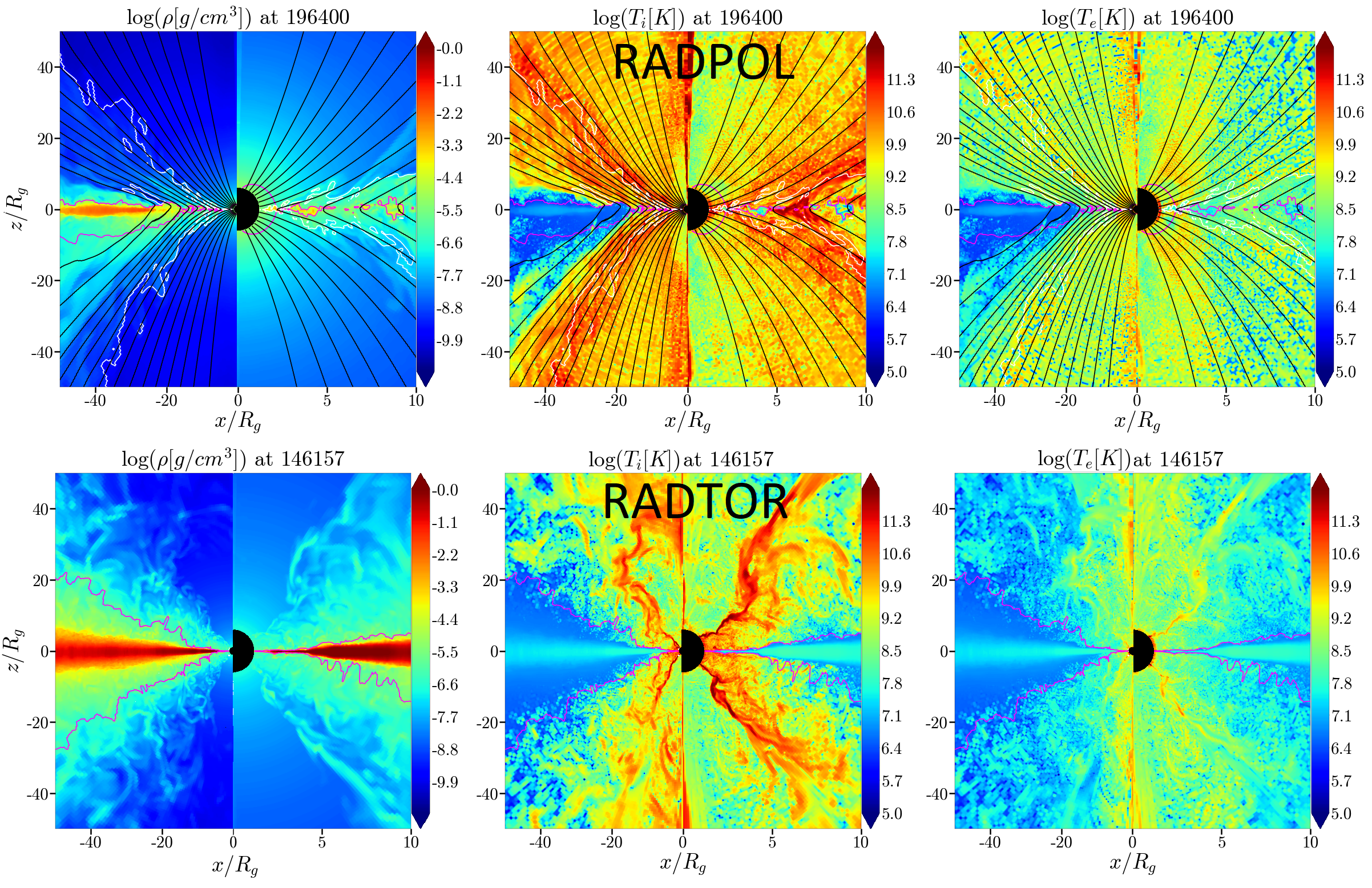}
\end{center}
\vspace{-14pt}
\caption{The presence of large-scale poloidal magnetic flux leads to the development of a two-phase medium: a low-density, thick, hot corona-like accretion flow with patches of cold gas floating through it (top row, model RADPOL) within $r \lesssim 20 r_g$. In contrast, in the absence of the poloidal magnetic flux the cold, thin flow extends down to the black hole (bottom row, model RADTOR). The three columns, from left to right, show vertical slices through the density, ion temperature, and electron temperature; the right hemispheres show zoom-ins on the left hemispheres. Magnetic field lines are shown in black, jet boundary in white (${p_{mag}}={\rho}c^2$), and the last scattering surface in pink ($\tau_{es}=1$, integrated in vertical direction).  In RADPOL the accretion disk transitions into a magnetic pressure supported corona within $r \lesssim 20 r_g$, which decouples into a 2-temperature plasma of hot, radiating, electrons and very hot, non-radiating, ions. Optically thick patches of cold gas, visible as high density regions with $\tau_{es} \gtrsim 1$ within $r \lesssim 25 r_g$, cover a significant fraction of the surface area in the corona near the midplane. In RADTOR the plasma in the disk remains strongly coupled and optically thick. Even above the photosphere, the plasma is much colder than in RADPOL.}
\label{fig:AMR}
\end{figure*}

To allow for the development of a two-phase medium we have implemented two-temperature thermodynamics in \hammer{}, and combined it with our M1 radiation scheme. In \hammer{}, we evolve the ion and electron entropy tracer, $\kappa={p_{e,i}}/{\rho^{\Gamma -1}}$, where $p_{e,i}$ is the electron and ion pressure, respectively, $\rho$ is the fluid-frame rest-mass density. In this work, we adopt $\Gamma = 5/3$ for both electrons and ions: we have verified that this is a safe choice by carrying out simulations with variable polytropic index for the electrons. The total dissipation is calculated as the difference between the internal energies computed from the total energy equation and from the entropy tracers \citep{Ressler2015}. For the purposes of partitioning the dissipated energy between the ions and electrons, we assume here that the dissipation ultimately occurs via magnetic reconnection in unresolved current sheets.  The dissipated energy is divided between the ions and electrons using an analytical prescription derived from particle-in-cell simulations \citep{Rowan2017}. We account for the energy exchange via Coulomb collisions between ions and electrons through an implicit source term \citep{Sadowski2017}. Due to the low radiative efficiency of ions as compared to electrons, when a two-temperature fluid forms, we assume that radiative cooling reduces the electron entropy and keeps the ion entropy constant.

\hammer{} allows broad flexibility in the choice of a coordinate system. Here, we adopt spherical polar coordinates, $r, \theta, \phi$, and choose a uniform grid in the $\log r$, $\theta$, and $\phi$ variables. We use a base resolution of $1020 \times 432 \times 288$ cells in the three dimensions, respectively. We then use AMR to increase this resolution in the regions of interest, as we describe below. We place the inner radial boundary inside of the event horizon (with $\ge5$ radial cells inside of the event horizon at the base level: this ensures that the inner radial grid boundary is properly causally-disconnected from the black hole exterior). We place the outer radial boundary at $R_{out}=10^4 r_g$. To resolve the structure of the thin equatorial disk, we quadruple the resolution for $5 r_g \lesssim r \lesssim 120 r_g $ within $7.5^{\circ}$ of the midplane to an effective resolution of $4080 \times 1728 \times 1152$ cells by adding 2 layers of static mesh refinement. In addition, we use 4 levels of local adaptive time-stepping, which adaptively sets the timestep based on the local Courant condition \citep{Liska2020}. This increases the timestep in the outer disk, which speeds up and improves the accuracy of the simulations \citep{Chatterjee2019}. To prevent cell squeezing near the poles ($\theta = 0, \pi$), we progressively reduce the $\phi$-resolution to $N_{\phi}=18$ within $30^{\circ}$ from each pole (we leave the resolutions in the other two dimensions unaffected). We use outflow boundary conditions in $r$-, transmissive boundary conditions in $\theta$- and periodic boundary conditions in $\phi$-directions. The grid features in total $\sim 0.45 \times 10^9$ cells and achieves a computational speed of effectively $1.5 \times 10^7$ zone-cycles/s/GPU (taking into account the speedup from the local adaptive timestepping). This is an order of magnitude less compared to the non-radiative version of \hammer{}. The grid is visualized in Fig.~\ref{fig:gridplot}.

\section{Physical Setup}
\label{sec:Physical_Setup}
To study the effect of large-scale magnetic flux on the structure of the inner disk, we consider two models. We initialize the model RADPOL with a purely poloidal magnetic field and model RADTOR with a purely toroidal magnetic field. Both models feature a rapidly spinning black hole of dimensionless spin $a=0.9375$ and use a $\Gamma=5/3$ for ions and electrons. Since ideal GRMHD is unable to describe the physical processes responsible for injecting gas in the jet funnel we use drift-frame density floors \citep{Ressler2017} to ensure that ${\rho c^2}\ge{p_{mag}}/12.5$ in the jets.

Model RADPOL starts with a thick equatorial torus in hydrostatic equilibrium with the inner radius at $r_{in}=12.5 r_g$, density maximum at $r_{max}=25 r_g$, and outer radius $r_{out}=200 r_g$ \citep{Fishbone1976}. The magnetic field has the vector potential $A_{\phi} \propto (\rho-0.05)^2 r^2$. We normalize this magnetic field such that $\beta=\max{ p_{gas}}/\max{p_{mag}}=30$ where $p_{gas}$ and $p_{mag}$ are the gas and magnetic pressures, and `$\max$' denotes the maximum taken over the torus. This setup is intended to represent the hard intermediate state, during which a hard-state thick torus collapses into a thin accretion disk due to radiative cooling (e.g. \citealt{Esin1997}). This radiative collapse is modelled by cooling the torus on the orbital timescale using a cooling function (see Sec.~\ref{sec:Results}). This cooling process is more thoroughly described in \citet{Liska2018C}, who used the same setup but ran the simulation at a lower resolution and for a shorter duration.

Model RADTOR starts with a thin equatorial accretion disk on Keplerian orbits in approximate hydrostatic equilibrium whose density distribution, $\rho(r,z) \propto r^{-1}\exp{(-{z^2}/{2h^2})}$, extends from an inner radius, $r_{in}=6 r_g$, to the outer radius, $r_{out}=76 r_g$ at the constant $h/r = 0.03$. Magnetic field is described by a vector potential $A_{\theta} \propto (\rho-0.0005)r^{2}$, which is normalized to give an approximately uniform $\beta \sim 7$. This setup is intended to represent the high soft state, where neither large-scale poloidal magnetic flux nor jets are present, presumably because all net poloidal magnetic flux has diffused out after the disk has settled into a geometrically thin disk (e.g. \citealt{Begelman2014}). 

\begin{figure*}
\begin{center}
\includegraphics[width=1.0\linewidth,trim=0mm 0mm 0mm 0,clip]{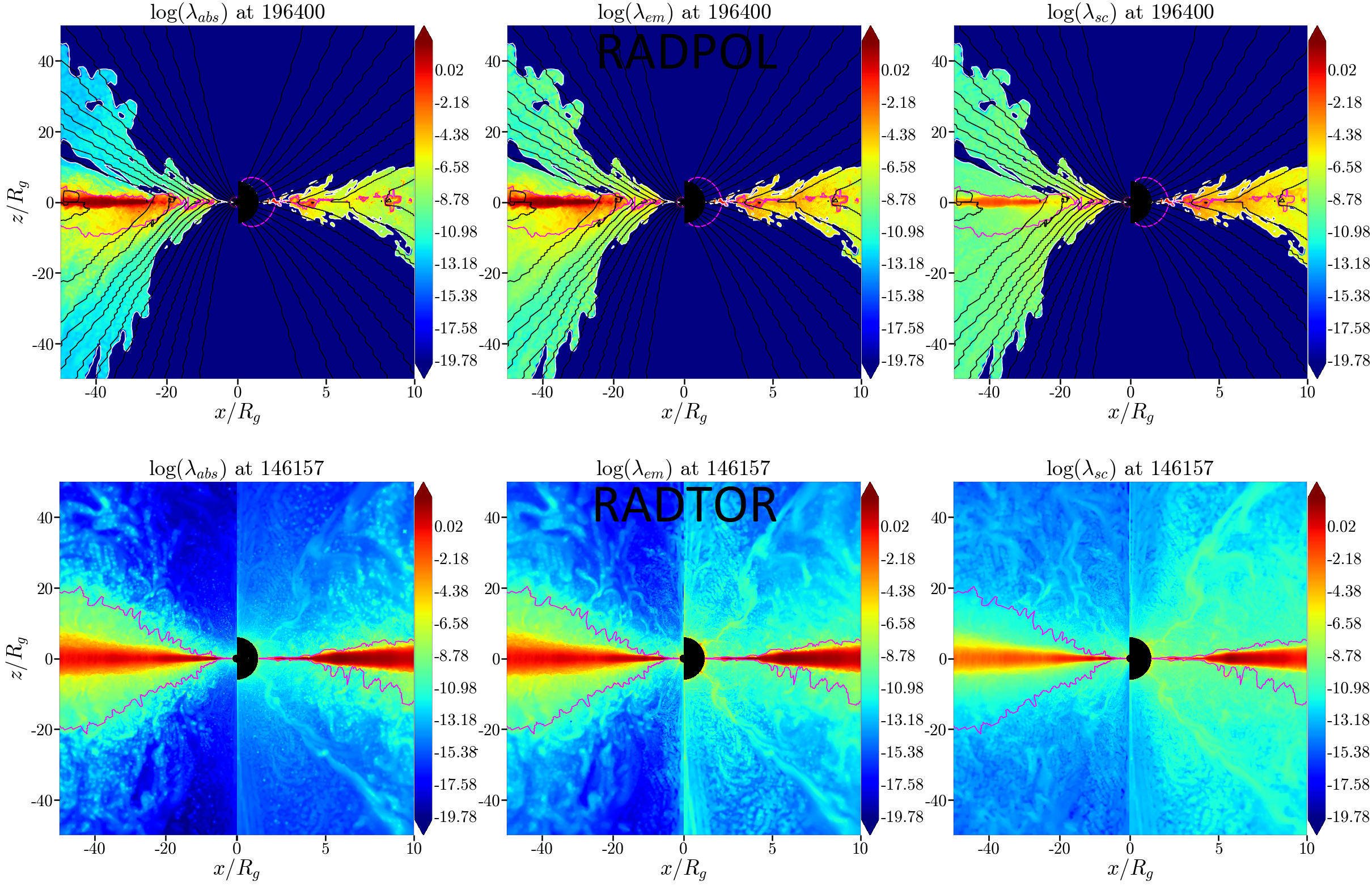}
\end{center}
\vspace{-14pt}
\caption{
The absorption heating rate ($\lambda_{abs}$, left column), emission cooling rate ($\lambda_{em}$, middle column), and Compton cooling rate ($\lambda_{sc}$, right column) for models RADPOL (top row) and RADTOR (bottom row). The right hemispheres show zoom-ins of the left hemisphere.  Magnetic field lines are shown in black, jet boundary in white (${p_{mag}}={\rho}c^2$), and the last scattering surface in pink ($\tau_{es}=1$, integrated in vertical direction). For clarity, we have removed data from the jet, whose thermodynamics cannot be modeled in GRMHD as explained in Sec.~\ref{sec:Results}. Cooling in the corona of RADPOL ($r \lesssim 25 r_g$) is concentrated near the equatorial plane, where the plasma has electron temperatures of $T_{e} \gtrsim 5 \times 10^8 K$. Here Compton cooling is expected to give rise to hard non-thermal emission that dominates over thermal emission. Since the plasma is optically thick (and self-absorbed) in model RADTOR, a more detailed ray-tracing analysis will be necessary to calculate the hardness of the spectrum.
}
\label{fig:compton}
\end{figure*}

%This gas injection is performed in the drift frame \citep{Ressler2017}, which does not lead to artificial drag on the field lines in the jet as in the ZAMO frame or to runaway energy gains as in the fluid frame.
\section{Results}
\label{sec:Results}
We evolve both models in two stages. During the first stage, we use neither the radiation transport nor two-temperature thermodynamics. Instead, we use a cooling function \citep{Noble2009} to cool the disk to a target thermal scaleheight of $({h}/{r})_{init}=0.03$: this cooling process takes $\sim 10^3 r_g/c$ for RADPOL and mimics the catastrophic cooling of a BHXRB disk undergoing an outburst as described in \citet{Liska2018C}. We initialize RADTOR at the target temperature profile corresponding to the same disk scaleheight, $({h}/{r})_{init}=0.03$. We evolve RADPOL for $t \sim 188,840 r_g/c$ and RADTOR for $t \sim 136,075 r_g/c$, which gives both models sufficient time to reach an approximately constant accretion rate on the BH. This allows us to select a well-defined accretion rate (with respect to the Eddington rate) before we turn on radiation.

The left panel in Figure~\ref{fig:init} shows that RADPOL accumulates large-scale poloidal magnetic flux in the inner disk: this is similar to the model described in \citet{Liska2018C}, which is essentially the same as RADPOL, apart from a small disk tilt, numerical grid, and duration. Model RADTOR does not generate nor advect any significant large scale poloidal magnetic flux as was suggested in previous work considering geometrically thick disks \citep{Liska2018B}. The reason for this is unclear, but it might involve the inability of geometrically thin disks to generate and/or advect magnetic flux inwards (e.g. \citealt{Lubow1994}). It is also possible that the simulation was not run long enough to capture this effect for geometrically thin disks, which have a 2 orders of magnitude longer viscous time than the disk presented in \citealt{Liska2018B}.  The flux accumulation in RADPOL is accompanied by a decrease in density at $r \lesssim 20r_g$ and the development of a magnetically arrested disk (MAD, \citealt{Tchekhovskoy2011}): the accretion in the inner disk proceeds through magnetic Rayleigh-Taylor instabilities (RTI), as is expected in a MAD (e.g. \citealt{Tchekhovskoy2011, Mckinney2012}). Neither of these two effects are observed for RADTOR (right panel of Fig.~\ref{fig:init}) and highlights the importance of the net poloidal magnetic flux for driving the disk truncation.

In the second stage, we set $M_{BH}=10 M_{\odot}$, renormalize the density such that $\langle \dot{M} \rangle \sim 0.35 \dot{M}_{Edd}$, and set $T_{rad}=T_{e}=T_{i}$, where we define $\dot M_{Edd}=L_{Edd}/\eta$ with $\eta = \eta_{NT}=0.178$ and $L_{Edd}={4 \pi GM_{BH}}/{\kappa_{es}}$. We assume solar abundances of hydrogen, helium and metals ($X=0.7, Y=0.28, Z=0.02$). We then run both simulations for another $t\sim 1.35 \times 10^4 r_g/c$ in full radiation-transport two-temperature GRMHD: for time interval $t\in[188840,203360] r_g/c$ in RADPOL and $t\in[136076,148997] r_g/c$ in RADTOR. We use blackbody Comptonization for both models, except during $t\in[191840,193650] r_g/c$ in RADPOL, where we use photon-number conserving Comptonization. In addition, we use for RADPOL during $t\in[193650,196550] r_g/c$ a lower density floor of ${\rho c^2}\ge{p_{mag}}/25$. Additionally, for the time interval of $t\in[188840,192630] r_g/c$ in RADPOL we limit the ratio of the electron to ion temperatures to $T_e/T_i \gtrsim 10^{-2}$ instead of the usual $T_e/T_i \gtrsim 10^{-4}$. These tests validate that the results presented in this article are not sensitive to the Comptonization routine employed or the value of the density floors as can be witnessed by the animations on our \href{https://www.youtube.com/playlist?list=PLDO1oeU33GwlkbTgY9gybClXqqv0ky7Ha}{YouTube channel} and in Figure~\ref{fig:tempb_tot}. However, setting the temperature floor to $T_e/T_i \gtrsim 0.01$ artificially increased the temperature of the electrons in current sheets near the midplane within $r \lesssim 5 r_g$ of the black hole. 

The top row in Figure~\ref{fig:AMR} (see also our \href{https://www.youtube.com/playlist?list=PLDO1oeU33GwlkbTgY9gybClXqqv0ky7Ha}{YouTube channel} for animations of the figure) shows that in RADPOL the inner disk decouples into a hot, optically thin, two-temperature plasma and can thus be associated with a corona. Since the physical processes such as pair creation (e.g. \citealt{Levinson2018}) responsible for mass loading of the jet cannot be modelled in GRMHD simulations, which artificially inject gas in the jet for the scheme to remain numerically stable, the jet thermodynamics fall beyond the scope of this work. Electrons in the corona reach temperatures of $T_{e} \gtrsim 5 \times 10^8 K$ while the ions reach temperatures of $T_{i} \gtrsim 10^{10} K$ irrespective of the Comptonization routine employed. Dissipation, most likely caused by magnetic reconnection in current sheets near the midplane (e.g. \citealt{RipperdaLiskaEtAl2021}), leads to localised heating of the plasma to temperatures of $T_i \sim 10^{12}K$ and $T_{e} \sim 10^9 K$, which suggest that electrons radiate their heat locally (see also \citealt{Beloborodov2017}). A significant part of this heated plasma flows out as a wind along the magnetic field lines (Fig.~\ref{fig:velplot}). On the other hand, pockets of cold gas survive and fall into the black hole predominantly along the equatorial plane. As explained in Sec. \ref{sec:conclusion} this has interesting consequences for reflection modelling.

Figure~\ref{fig:compton} shows the cooling rate of the plasma due to emission ($\lambda_{em}=\rho \kappa_{em} a_{rad} T_{e}^{4}$) and inverse Compton scattering ($\lambda_{sc}=\rho \kappa_{es} E_{rad}\, {4 k_{b} T_{e}}/{m_{e} c^{2}}$), in addition to the heating rate through absorption ($\lambda_{abs}=\rho \kappa_{abs} E_{rad}$). As can be seen, the cooling is predominantly concentrated near the equatorial plane which has an optial depth of order unity ($\tau_{es}\sim1$). Since the Compton scattering rate exceeds the emission rate, it is conceivable that most radiation actually gets Compton up-scattered before leaving the corona. We estimate that the Compton-$y$ parameter ($y=\tau_{es}\,{4 k_{b} T_{e}}/{m_{e}c^{2}} $), which measures the average fractional energy gain of photons due to Comptonization (e.g. \citealt{Rybicki1986}), can easily exceed $y \gtrsim 1$ in the $T_e \gtrsim 5 \times 10^8 K$ plasma near the midplane of the corona. Pinpointing the exact value of the Compton y-parameter is non-trivial since absorption will play an important role in some optically thick patches near the midplane. A more detailed calculation that takes into account emission, absorption and scattering in the plasma along photon geodesics will be required to further constrain the Compton-$y$ parameter. Nevertheless, this work suggests that the corona will up-scatter and significantly harden thermal radiation from the disk and cyclo-Synchrotron emission from the corona (see also \citealt{Dexter2021, Scepi2021}). In contrast, the bottom row in Figure~\ref{fig:AMR} shows that in RADTOR the ions and electrons are strongly coupled in the cold disk, which remains optically thick to both scattering and absorption. Since this disk extends all the way to the event horizon, we expect a strong thermal emission component in the emergent spectrum. Compton up-scattering of disk photons by the winds sandwiching the inner disk to temperatures necessary to significantly harden the spectrum is not expected since these winds are predominantly cold ($T_{e} \lesssim 5 \times 10^{7} K$). The hot pockets of gas with temperatures $T_e\gtrsim 10^8 K$ within $r \lesssim 5 r_g$ in figure \ref{fig:AMR} are relatively rare and seem to be more prominent at later times when the disk becomes numerically under-resolved due to thermal collapse (Sec. \ref{sec:Ap_convergence}).

\begin{figure}
\centering
\includegraphics[clip,trim=0.0cm 0.0cm 0.0cm 0.0cm,width=\columnwidth]{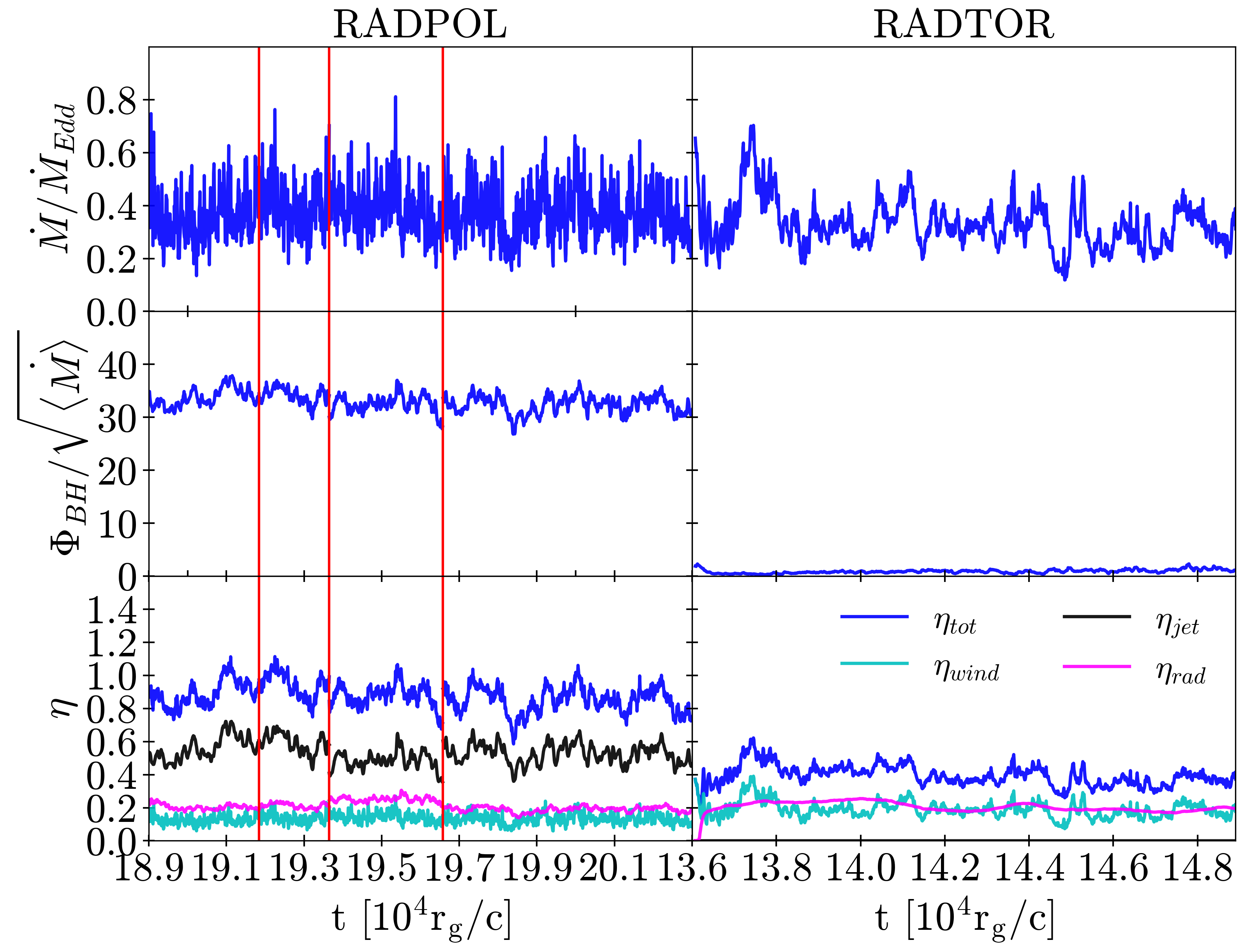}
 \vspace{-10pt}
\caption{The first demonstration of highly efficient energy extraction from a rapidly spinning black hole by a radiatively efficient accretion flow in a radiation GRMHD simulation. From top to bottom, the panels show the accretion rate in Eddington units, normalized magnetic flux on black hole and outflow efficiencies (total in blue, jet in black, wind in cyan, and radiative in purple; see the legend) for the radiative stage of the simulations. Both RADPOL and RADTOR accrete at $\dot{M}/\dot{M}_{Edd} \sim 0.35$. The red lines demarcate the time intervals when we use either photon-conserving Comptonization  ($t\in[191840,193650] r_g/c$) or a higher density floor ($t\in[193650,196550] r_g/c$). While RADPOL floods the black hole with the large-scale poloidal magnetic flux and reaches the MAD state, RADTOR does not develop any net large-scale poloidal magnetic flux. In RADPOL, the total efficiency exceeds $\eta_{tot}\sim90\%$, for the first time demonstrating highly efficient energy extraction from a radiatively efficient accretion flow out of spinning black hole in a radiation GRMHD simulation. The efficiencies of the jet and winds reach respectively $\eta_{jet} \sim 50\%$ and $\eta_{wind} \sim 15\%$. In contrast, in the RADTOR simulation, the total efficiency is much smaller, $\eta_{tot} \sim 40\%$, with a much smaller $\eta_{jet} \lesssim 1\%$, and and approximately the same $\eta_{wind} \sim 20\%$. Both models achieve similar radiative outflow efficiencies, $\eta_{rad} \sim 20\%$.}
\label{fig:timeplot}
\end{figure}

We show the accretion rate $\dot{M}=-\int_{0}^{2\pi}\int_{0}^{\pi} \rho u^{r} \sqrt{-g} d \theta d \phi$, with the integral evaluated at $r = 5 r_g$ to avoid the contamination by the density floors, and dimensionless black hole poloidal magnetic flux $\phibh=0.5{\int_{0}^{2\pi}\int_{0}^{\pi} |B^{r}| \sqrt{-g} d \theta d \phi}/{\langle \dot{M} \rangle}^{1/2}$,  with the integral evaluated at the event horizon, $r = r_{H} \simeq 1.3 r_g$, in the upper and middle panels of Figure~\ref{fig:timeplot}.  In the RADPOL model, the accretion disk transports so much of the initial poloidal magnetic flux to the black hole that it becomes as strong as gravity, obstructs the accretion, reaches a saturated value, and leads to a MAD. The normalized magnetic flux $\phi_{BH} \sim 33$ is about two-thirds of that in thick accretion disks \citep{Tchekhovskoy2011} and consistent with the MAD saturation value in thin accretion disks (Cheng, Liska et al 2022, in prep). The lower panel of Figure~\ref{fig:timeplot} shows that RADPOL launches powerful jets and winds. Namely, we show the jet ($\eta_{jet}$) and wind ($\eta_{wind}$) energy efficiencies, which we obtain by normalizing the electromagnetic plus fluid energy fluxes at $r = 5 r_g$ in the jets and winds by the mass accretion rate (see \citealt{Liska2018C}). We define the jet-wind boundary by the ${p_{mag}}={\rho}c^2$ criterion. In RADPOL, we find $\eta_{wind} \sim 15\%$ and $\eta_{jet} \sim 50\%$. In RADTOR, we find $\eta_{wind} \sim 20\%$ and $\eta_{jet} \lesssim 1\%$. In both models the associated winds are significantly hotter than the disk. We compute the radiative efficiency from the normalized radiation energy flux measured at $r = 100 r_g$ and find $\eta_{rad} \sim 20 \%$ in both models. This is remarkably close to the \citet{Novikov1973} value $\eta_{NT} \sim 17.8\%$. 

\begin{figure}
    \centering
    \includegraphics[clip,trim=0.0cm 0.0cm 0.0cm 0.0cm,width=\columnwidth]{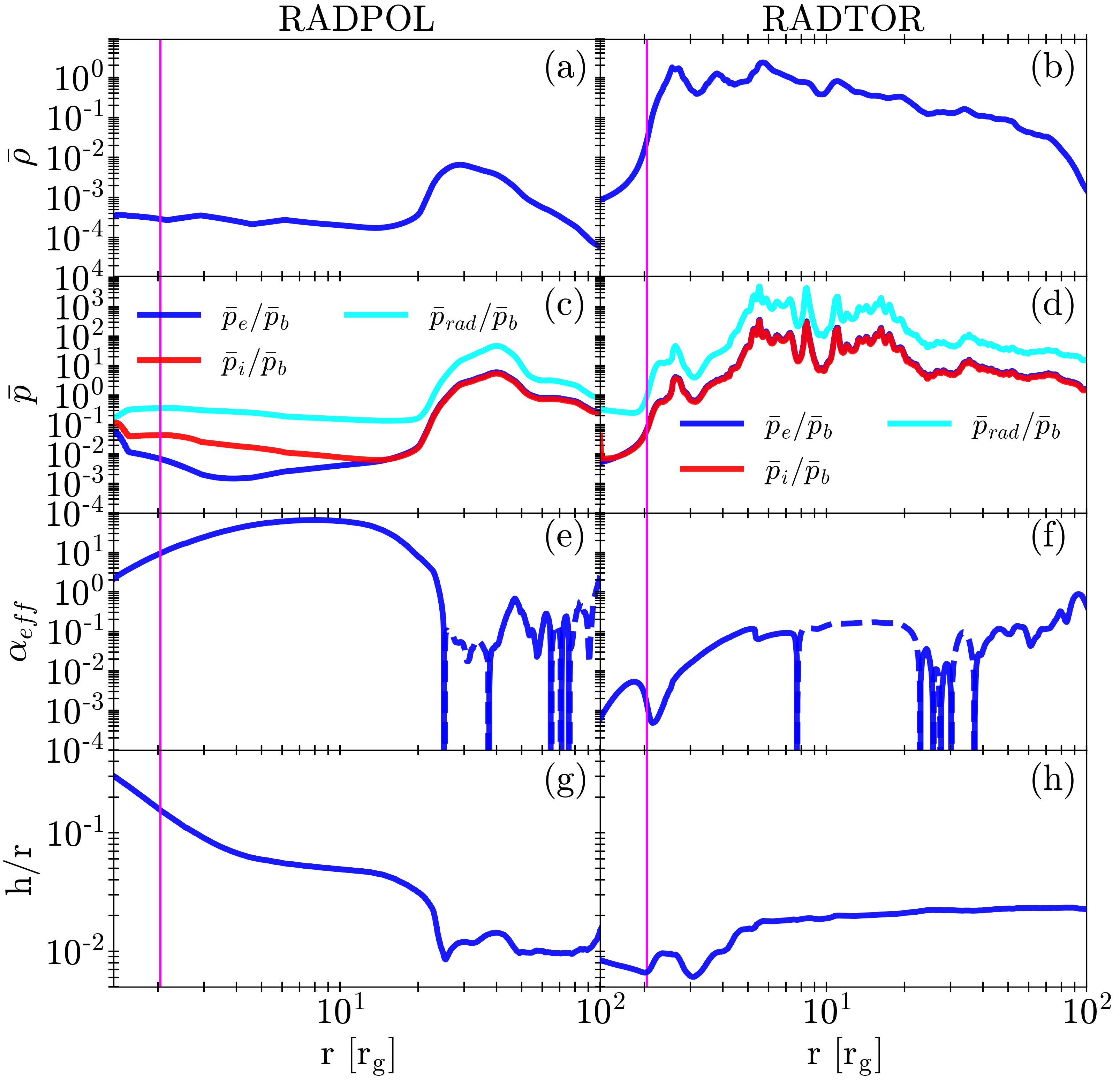}
    \vspace{-10pt}
    \caption{Low-density, magnetically-dominated, rapidly accreting, thicker corona--like accretion flow develops near the black hole only in the presence of large-scale poloidal magnetic flux. From top to bottom, the panels show density-averaged midplane density $\bar{\rho} =\int \rho^2 dV/\int \rho dV$ (panels a and b); the ratios of density-averaged pressures, $\bar{p}=\int \rho p dV/\int \rho dV$ (panels c and d); effective $\alpha$-viscosity (panels e and f; dashed lines indicate negative values); and scaleheights (panels g and h) time-averaged over the radiative evolution stage of the simulations. The left panels show RADPOL and right panels RADTOR results. The vertical magenta lines show the location of the ISCO. While in both models the outer disk ($r \gtrsim 20r_g$) is radiation pressure dominated, the inner disk in RADPOL becomes magnetic pressure dominated and forms a two-temperature plasma (with $\bar p_e\ll \bar p_i$). This transition to a magnetic pressure supported corona in RADPOL is accompanied by a sharp drop in density and an increase in the effective scaleheight. The presence of large scale poloidal magnetic flux leads to rapid accretion in RADPOL as evidenced by the extremely high value of $\alpha_{eff}$.}
    \label{fig:radplot}
\end{figure}

As Figure~\ref{fig:AMR} shows, in RADPOL the excess magnetic flux remains in the inner disk, $r \lesssim r_{\rm tr}\approx20 r_g$. The left column in Figure~\ref{fig:radplot} shows that this leads to the formation of a magnetic pressure supported low-density inner disk coupled to a radiation pressure supported high-density outer disk. The transition between these two regimes occurs at a well-defined `magnetic truncation' radius, $r_{\rm tr}$, which is set by the distance out to which the inner disk is flooded by the poloidal magnetic flux. Similar to the single-temperature simulation stage (Fig.~\ref{fig:init}), here the accretion in the inner disk also proceeds through magnetic RTI and is the radiative analog of the MAD (see also \citealt{Teixeira2017}). On the other hand, the disk in RADTOR remains radiation pressure dominated and does not contain any large-scale poloidal magnetic flux. We define the effective viscosity as $\alpha_{eff}=-\langle{v_{r}v_{\phi}\rangle_{\rho}}/{\langle ({h}/{r})_{init}^{2} v_{\phi}^2\rangle_{\rho}}$, where $\langle \dots\rangle_{\rho}$ is the density-weighted angular average, and $v_r$ and $v_\phi$ are the physical radial and azimuthal velocity components, respectively.  Figure~\ref{fig:radplot}(g,h) shows that $\alpha_{eff} \sim 10^{1}$ in RADPOL and $\alpha_{eff} \sim 10^{-2}$ in RADTOR. This illustrates that the presence of large scale magnetic flux, as in RADPOL,  leads to more rapid accretion of gas (e.g. \citealt{Ferreira2006}). Please note that even when $\alpha_{eff} \sim 100$ the ratio between the radial and azimuthal velocities becomes $v_{r}/v_{\phi}=(h/r)^{2} \alpha \lesssim 0.1$, suggesting the infalling gas has a significant toroidal velocity component capable of producing a broadened iron line.

In both models the scaleheight of the radiation pressure supported part of the disk shrinks to a density averaged scaleheight ${h}/{r}= \langle|\theta -\pi/2|\rangle_{\rho} \sim 0.015-0.02$ (Fig.~\ref{fig:radplot} panels g and h and \href{https://www.youtube.com/playlist?list=PLDO1oeU33GwlkbTgY9gybClXqqv0ky7Ha}{YouTube channel}) at which point (when the disk scaleheight or MRI wavelength become resolved by less than $\sim 8$ cells beyond $r \gtrsim 5 r_g$) we terminate the simulation. This runaway cooling effect suggests that both disks are subject to viscous and/or thermal instabilities \citep{Lightman1974, Shakura1976}. However, in RADPOL magnetic pressure stabilizes the inner disk against thermal collapse (see also \citealt{Sadowski2016_mag, Jiang2019}), which settles into a scaleheight of ${h}/{r} \sim 0.05-0.2$. Please note that the outer disk in both RADPOL and RADTOR is radiation pressure dominated and thus remains thermally unstable. Future simulations with higher resolutions and longer runtimes will investigate whether magnetic pressure or other physical processes eventually stabilize the entire accretion disk against runaway collapse (e.g. \citealt{Begelman2007}). This was supported by radiative GRMHD simulations of thin accretion disks seeded with equipartition strength magnetic fields (e.g. \citealt{Sadowski2016_mag, Lancova2019}).

\section{Discussion and Conclusion}
\label{sec:conclusion}
In this work we have presented the first radiation-transport two-temperature GRMHD simulations of luminous sub-Eddington accretion disks that accrete close to the Eddington limit ($L/L_{edd} \gtrsim 0.01$). Previous work in GRMHD only addressed highly sub-Eddington disks with  $\lambda = \langle \dot{M}\rangle/\dot{M}_{Edd} \ll 0.01$ (e.g. \citealt{Sadowski2017, Chael2017, Chael2019, Dexter2021}). More specifically, the simulations in this article accrete at the Eddington ratio $\lambda = \langle \dot{M}\rangle/\dot{M}_{Edd} \sim 0.35$. One of the simulations, model RADTOR, had no net poloidal magnetic flux, while the other one, model RADPOL, became saturated with poloidal magnetic flux and entered a MAD state. These models were chosen to demonstrate the importance of magnetic flux saturation on the (thermo-)dynamics of the plasma and do not attempt to describe a possible transition between RADPOL and RADTOR and/or to rule out other physical mechanisms that can lead to spectral hardening or disk truncation (e.g. \citealt{Meyer1994}).  We demonstrated that the saturation of magnetic flux in the inner disk of model RADPOL creates a sharp transition between a radiation pressure supported outer disk and magnetic pressure supported inner corona. The sharp transition between the two occurs at a magnetic truncation radius of $r_{\rm tr} \sim 20 r_g$. Powerful jets, winds, and radiative outflows with a combined efficiency of $\eta_{tot} \gtrsim 90\%$ emerge in the presence of this poloidal magnetic flux. All of these characteristics are consistent with accretion in the hard (-intermediate state) of BHXRBs. 

The corona is best described by a MAD where radial magnetic pressure gradients prevent ordered quasi-axisymmetric accretion of gas (e.g. \citealt{Narayan2003, Igumenschev2003, Igumenshchev2008, Tchekhovskoy2011, Mckinney2012, Begelman2021}). Accretion of gas in the corona proceeds through magnetic RTI instabilities originating at the truncation radius. While a significant fraction of the infalling gas gets heated by magnetic reconnection to $T_i \gtrsim 10^{10}K$ and $T_e \gtrsim 5 \times 10^{8}K$, we also observe cold, optically thick, clumps of gas falling into the black hole (Fig.\ref{fig:AMR}). This could potentially explain broad iron line emission observed in the hard spectral states of XRBs and AGN (e.g. \citealt{Reis2010}) without the need for the corona to recondense into a thin accretion disk near the black hole (e.g. \citealt{Meyer-Hofmeister2011}).

Our model differs substantially from most truncated accretion disk models in the literature (e.g. \citealt{Esin1997, Ferreira2006, Begelman2014}). In such models heat conducted from the corona into the upper layers of the disk leads to the disk's gradual evaporation into a corona (e.g. \citealt{Meyer1994, liu98, Qian2007}). However, in magnetically truncated disks the transition from a disk into a corona occurs at a well-defined radius which is determined by the location of the magnetospheric radius (e.g. \citealt{Igumenshchev2009}). Since accretion within the magnetospheric radius proceeds through RTI modes, we expect that the power spectrum of the Comptonized emission from the corona will differ substantially from the outer disk's blackbody emission. In future work we will address how the cold-phase gas transitions into hot-phase gas in the corona. At the moment of writing, we suspect that magnetic reconnection in equatorial current sheets heats a small fraction of the cold-phase gas to extremely high temperatures, while leaving most of the cold-phase gas unaffected. The hot-phase case subsequently flows out along magnetic field lines (Sec. \ref{sec:Ap_outflows}). This is a conceivable scenario in a magnetic pressure supported plasma, but will require further analysis (Liska et al 2022C in prep).

Spectral state transitions (e.g., \citealt{Fender2004}) can potentially be explained by the outwards diffusion of magnetic flux (e.g. \citealt{Lubow1994, Begelman2014}). From an observational perspective, the truncation radius needs to move inwards to explain the softening of the spectrum during a hard-to-soft state transition (e.g. \citealt{Esin1997}) while the magnetic flux needs to diffuse out to explain the lack of any jets in the high-soft state (e.g. \citealt{Fender2004}). This has never been observed in GRMHD simulations since magnetic flux diffusion is expected to occur on much longer timescales than can be captured by state-of-the-art GRMHD models. Nevertheless, it is conceivable that outwards diffusion of the magnetic flux in our model will progressively reduce the magnetic truncation radius, which will cause the spectrum to become softer. However, even when the entire disk leaves the MAD state, residual large scale poloidal magnetic flux might still be present and launch powerful winds from the inner disk. Such winds could be significantly hotter compared to the winds in RADTOR and could potentially explain the hard emission tail observed in the soft (-intermediate) states of BHXRBs. In addition, other processes such as shocks (\citealt{Musoke2022}, Liska et al 2022B in prep) and emission from jets (e.g. \citealt{Markoff2005}) can lead to significant hardening of the spectrum.

Interestingly, \citet{Kinch2021} used an iterative post-processing approach \citep{Kinch2019} to generate spectra of thin accretion disks, which were simulated in GRMHD using a cooling function \citep{Noble2009, Noble2011}. In these simulations the complex radiative processes responsible for cooling the disk and corona (defined as gas outside of the $\tau_{es} \lesssim 1$ surface) were approximated with cooling functions \citep{Noble2009, Kinch2020}. The temperature of the radiation emitting electrons in the corona was determined by matching the cooling rate in the corona to the Compton emission rate of the electrons. Making a direct comparison between \citet{Kinch2021} and RADPOL/RADTOR simulations is difficult since both the initial configuration and numerical methods differ substantially. The thin disk in \citet{Kinch2021} is, for example, much thicker ($h/r \sim 0.05-0.06$) and contains a significant amount of poloidal magnetic flux, but stays below the MAD saturation value, suggesting neither RADPOL or RADTOR can describe it. In addition, our radiative GRMHD simulations model the radiative processes in the disk and corona such as emission, absorption, and scattering self-consistently. We also account for thermal decoupling between ions and electrons in the corona, and include feedback between radiation and gas. For example, in our radiative models the local dissipation rate is not guaranteed to match the local emission rate which can lead to runaway thermal collapse of the accretion disk (Sec.~\ref{sec:Results}).

By ray-tracing our simulation data (e.g \citealt{Krawczynski_Henric2021}) we will be able place unique constraints on the relative contributions of the cold-phase and hot-phase gas. However, generating self-consistent spectra will face several challenges. For example, magnetic reconnection in collisionless plasma leads to the formation of non-thermal particle populations (e.g \citealt{Sironi_2014, Beloborodov2017, Sironi2020, Navin2021}). Since ideal GRMHD is unable to model the full energy distribution of such particles, deviation from the thermal spectrum is a free parameter that should be quantified by future particle-in-cell (PIC) simulations. In addition, it is likely that there are some inaccuracies in the electron temperatures presented in this work. These stem from the intrinsic challenges associated with quantifying the dissipation rate in a highly magnetized plasma, and from the inherent limitations related to our two-moment (M1) radiation scheme which treats radiation as a highly collisional fluid. This works best in regions of high optical depth where emission and scattering of photons is localized. However, in the optically thin corona there are multiple radiation fields which M1 cannot capture. These include thermal radiation from the cold-phase gas and disk, Compton-scattered radiation from the hot-phase gas in the corona, and reflected coronal radiation from the cold-phase gas and disk. This can lead to irradiation of the dense gas clumps in the corona by each other and the hot-phase gas, which can potentially disperse the plasma to a less clumpy state than predicted by M1. In other words, future work will be necessary to determine if the clumpy structure in the corona survives when properly accounting for all radiation fields, and, if the irradiated clumps will indeed be able to produce a relativistic broadened iron line (which will require an elaborate calculation to determine the ionisation state of the plasma).

\section{Acknowledgements}
We thank Michiel van der Klis, Adam Ingram, Sera Markoff, Ramesh Narayan, and Eliot Quataert for insightful discussions.
An award of computer time was provided by the Innovative and Novel Computational Impact on Theory and Experiment (INCITE) program under award PHY129. This research used resources of the Oak Ridge Leadership Computing Facility, which is a DOE Office of Science User Facility supported under Contract DE-AC05-00OR22725. This research also used HPC and visualization resources provided by
the Texas Advanced Computing Center (TACC) at The University
of Texas at Austin, which contributed to our results via the LRAC allocation AST20011 and through Texascale Days (http://www.tacc.
utexas.edu). We acknowledge PRACE for awarding us access to JUWELS Booster at GCS@JSC, Germany. ML was supported by the John Harvard Distinguished Science Fellowship, GM is supported by a Netherlands Research School for Astronomy (NOVA), Virtual Institute of Accretion (VIA) postdoctoral fellowship and AT by the National Science Foundation grants AST-2009884, AST-1815304 and AST-1911080.

\section{Appendix}
\label{sec:Appendix}

\subsection{Formation of outflows}
\label{sec:Ap_outflows}
\begin{figure*}
    \centering
    \includegraphics[width=1.0\linewidth,trim=0mm 0mm 0mm 0,clip]{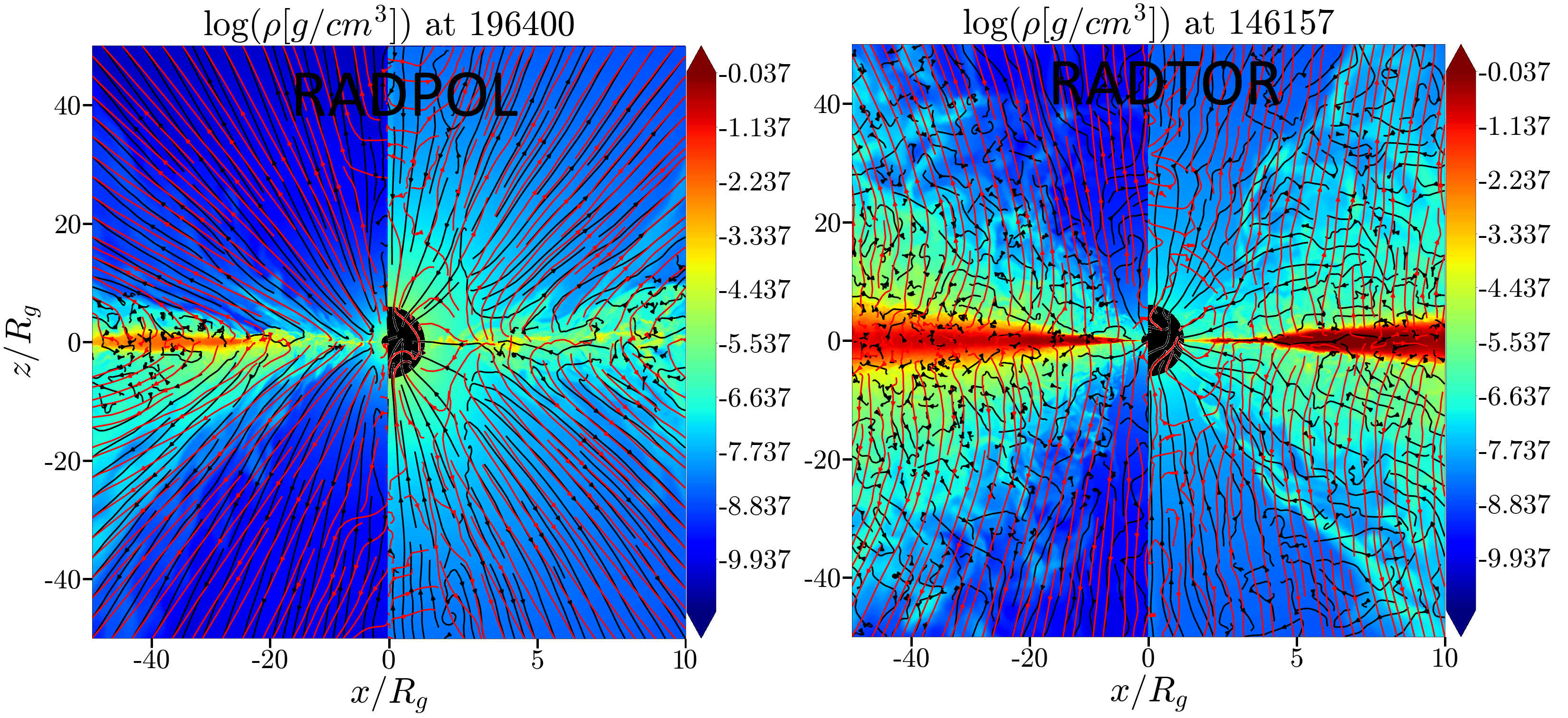}
    \vspace{-10pt}
    \caption{A transverse slice through density in RADPOL and RADTOR. The right hemisphere is a zoom-in of the left hemisphere. Velocity streamlines are illustrated in black and radiation streamlines are illustrated in red. In model RADPOL both the radiation and gas flow out radially. In RADTOR the gas velocity shows a chaotic structure, while the radiation escapes mostly vertical.}
    \label{fig:velplot}
\end{figure*}

Both RADPOL and RADTOR form radiative and mechanically driven outflows. In figure \ref{fig:velplot} we use streamlines to illustrate the direction of the radiation velocity (red) and gas velocity (black). Both of these outflows move out radially or vertically from the disk's midplane. In future work we plan to address the mass and energy ejection rates of these outflows and if these outflows indeed become gravitational unbound. Note that radiation is treated as a single fluid in the M1 approximation and thus only the net radiative energy flow is captured. In reality, the radiation field would consist out of a blackbody-component originating in the disk and Compton-scattered radiation originating from the corona which can be reflected of the disk. 

\subsection{Convergence}
\label{sec:Ap_convergence}
\begin{figure*}
\begin{center}
\includegraphics[width=1.0\linewidth,trim=0mm 0mm 0mm 0,clip]{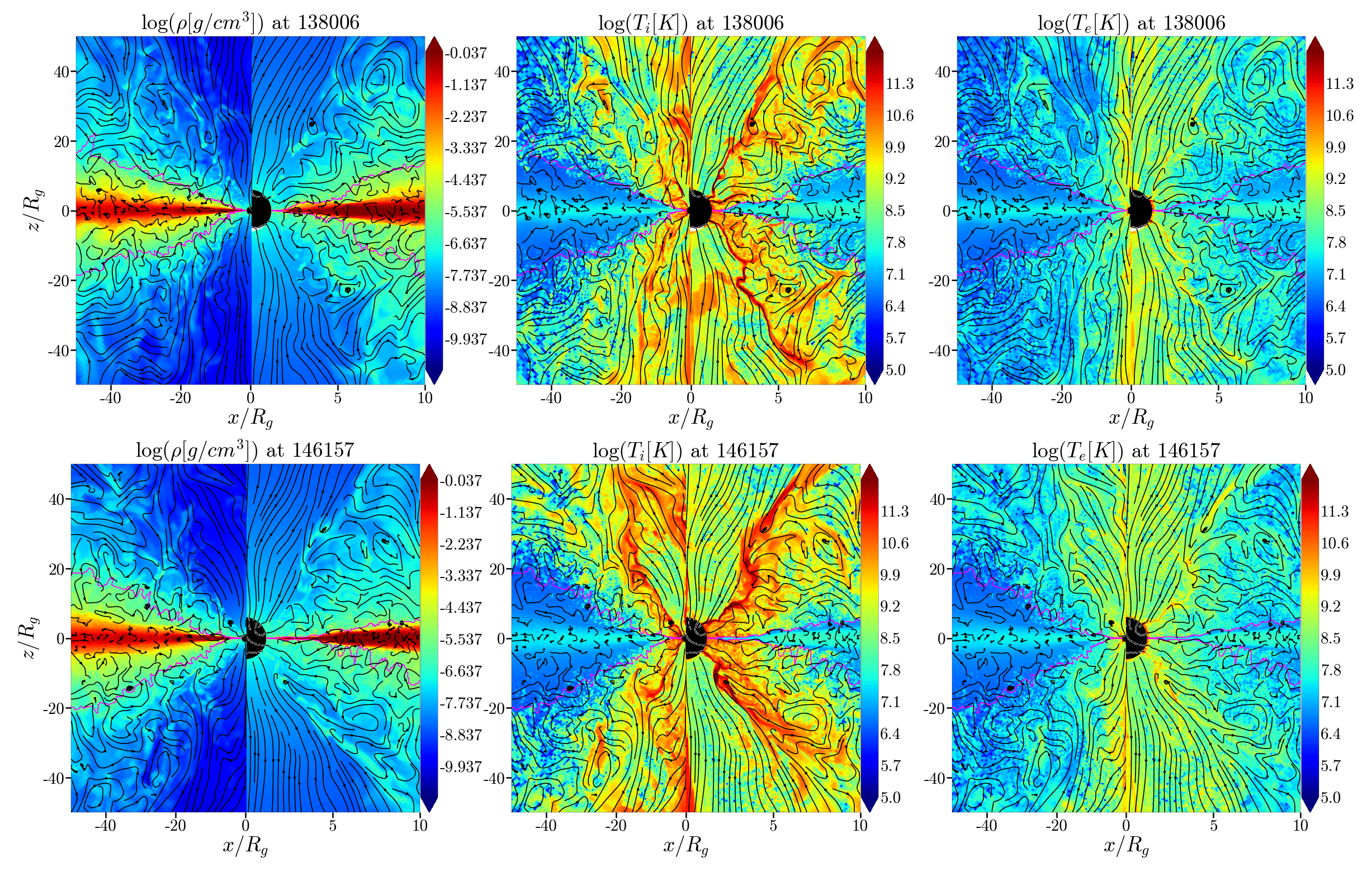}
\end{center}
\vspace{-14pt}
\caption{A transverse slice through density (left panels), ion temperature (middle panels) and electrons temperature (right panels) in RADTOR at early times (upper panels) and late times (lower panels). The right hemisphere is a zoom-in of the left hemisphere.  Magnetic field lines are visualised in black and the last scattering surface is pink. At late times the inner disk is subject to runaway thermal collapse, which causes the temperature of the gas above the photosphere (cyan lines) to modestly increase. This could be caused by insufficient resolution in the disk.}
\label{fig:collapseplot}
\end{figure*}

\begin{figure}
    \centering
    \includegraphics[clip,trim=0.0cm 0.0cm 0.0cm 0.0cm,width=\columnwidth]{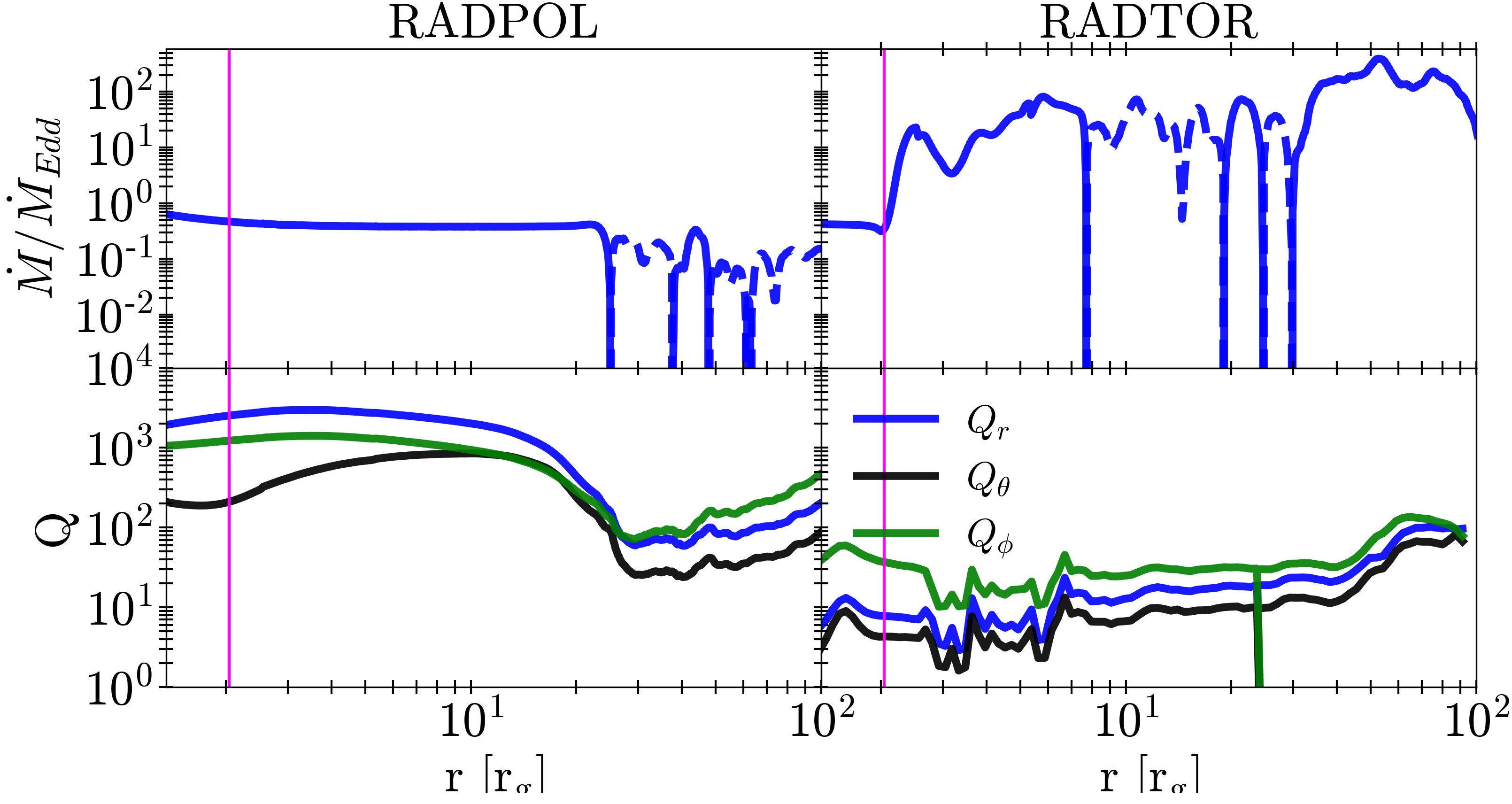}
    \vspace{-10pt}
    \caption{Accretion rate $\dot{M}$ (upper panels) and MRI quality factors $Q_{r, \theta, \phi}$ (lower panels) as function of radius averaged over the final $1000 r_g/c$ runtime of the respective simulation. Dashed lines show negative values. While RADPOL achieves inflow equilibrium within $r \lesssim 25 r_g$, RADTOR is thermally unstable and never achieves inflow equilibrium. The MRI quality factors give the number of cells per MRI wavelength. RADPOL is sufficiently resolved throughout the domain while RADTOR becomes underresolved ($Q \lesssim 10$) within $r \sim 10 r_g$ as the disk undergoes runaway thermal collapse.}
    \label{fig:Qplot}
\end{figure}

To quantify the convergence of our models we plot in Figure~\ref{fig:Qplot} the radial dependence of the mass accretion rate and MRI quality factors. In model RADPOL the mass accretion rate reaches a steady value within $r \lesssim 25 r_g$. This suggests RADPOL achieved inflow equilibrium in the MAD region of the disk and corona. In RADTOR no inflow equilibrium can be achieved since the disk collapses into an infinitely thin slab due to radiative cooling. The MRI quality factors ($Q_{r}$, $Q_{\theta}$, $Q_{\phi}$) are defined as the number of cells per MRI-wavelength ($\lambda_{MRI}$) weighted by $b^2 \rho$ ($Q_{r, \theta. \phi}=|\lambda_{MRI}|_{b^2 \rho}/N_{r, \theta, \phi}$). Resolving the MRI wavelength by at least 10 cells, and ideally by more than 20 cells(e.g. \citealt{Shiokawa2012, Porth2019}), is necessary to capture the the growth of the longest-wavelength MRI modes in the $\theta-$ and $\phi-$ direction. Both RADPOL and RADTOR have $Q_{\phi} \gtrsim 20$ throughout the domain. However, in RADTOR $Q_r$ and $Q_{\theta}$ reach a marginal $Q_r \sim Q_{\theta} \sim 6-10$ for $r \lesssim 10$. Since the initial setup only contains a toroidal magnetic field it is not unexpected that the poloidal components of the magnetic field become challenging to resolve, especially after the inner disk has undergone thermal collapse. We suspect that this can potentially heat up the plasma above the photosphere in RADTOR (Fig.~\ref{fig:collapseplot}).

\subsection{Effect of Comptonization and density floors}
\begin{figure*}
\begin{center}
\includegraphics[width=1.0\linewidth,trim=0mm 0mm 0mm 0,clip]{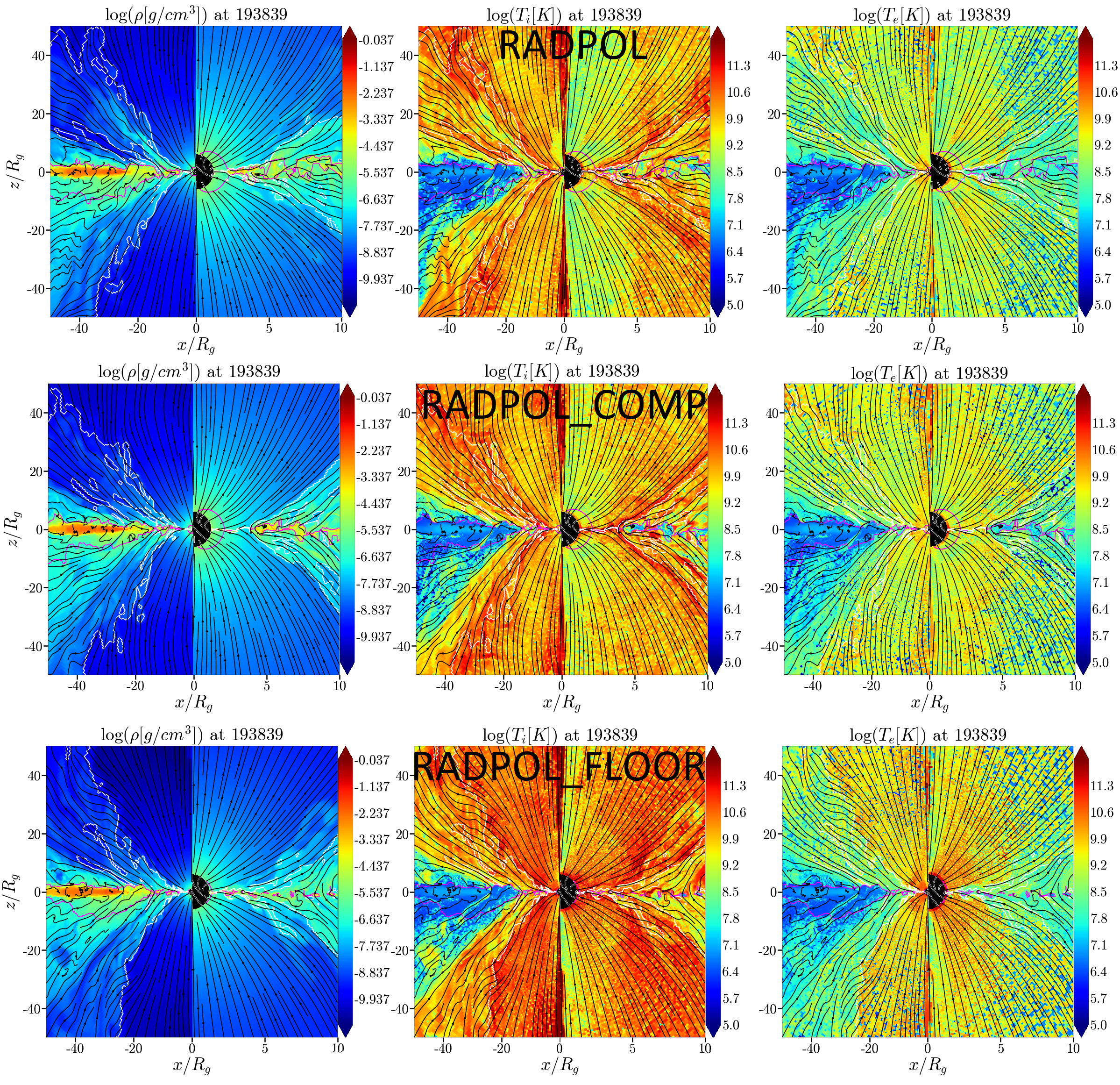}
\end{center}
\vspace{-14pt}
\caption{A transverse slice through density (left panels), ion temperature (middle panels) and electron temperature (right panel) for model RADPOL with photon non-conserving comptonization (RADPOL, top panels), with photon-conserving comptonization (RADPOL-COMP, middle panels) and with lower density floors (RADPOL-FLOOR, lower panels). The right hemisphere is a zoom-in of the left hemisphere. Magnetic field lines are black while the location of the jet-disk boundary is white, and the $\tau_{es}=1$ surface is pink. Based on the animated version of this figure (see supplementary materials), we can conclude that the temperature of the disk and corona remains relatively insensitive to these choices in the numerics. }
\label{fig:tempb_tot}
\end{figure*}

To quantify the effects of different treatments for the Comptonization and a lower density floor, we plot in figure \ref{fig:tempb_tot} a snapshot of each simulation, which was restarted at $t=188839 r_g/c$. From these figures and the animations in the supplementary materials we conclude there is no strong dependence on either of these two numerical choices for the temperature of the disk and/or corona supporting the conclusions presented in this paper. However, the temperature of the jet strongly depends on the value of the density floors. This is an inherent limitation of GRMHD and will need to be addressed by particle-in-cell simulations (e.g. \citealt{Levinson2018}).

\bibliography{new.bib}{}
\bibliographystyle{aasjournal}

%% This command is needed to show the entire author+affiliation list when
%% the collaboration and author truncation commands are used.  It has to
%% go at the end of the manuscript.
%\allauthors

%% Include this line if you are using the \added, \replaced, \deleted
%% commands to see a summary list of all changes at the end of the article.
%\listofchanges

\end{document}